\documentclass[ reprint, amssymb, amsmath, pre,cha]{revtex4-1}

\usepackage[colorlinks=true,linkcolor=blue]{hyperref}%
\usepackage{graphicx}
\usepackage{amsmath}
\usepackage{booktabs}

\newcommand{\la}{\ensuremath{\langle}}
\newcommand{\ra}{\ensuremath{\rangle}}

\newcommand{\vc}{\mathbf}

\usepackage{color}
\usepackage{amsmath,amssymb}%,wasysym}
\usepackage{graphicx}
\usepackage{empheq}
\usepackage{mathrsfs}

\begin{document}

\title{Testing thermal conductivity models with equilibrium molecular dynamics simulations of the one component plasma}
\author{Brett Scheiner }
\affiliation{Department of Physics and Astronomy, University of Iowa, Iowa City, Iowa 52240}

\author{Scott D. Baalrud}
\affiliation{Department of Physics and Astronomy, University of Iowa, Iowa City, Iowa 52240}

\begin{abstract}

Equilibrium molecular dynamics simulations are used to calculate the thermal conductivity of the one component plasma (OCP) via the Green-Kubo formalism over a broad range of Coulomb coupling strength, $0.1\le\Gamma\le180$. These simulations address previous discrepancies between computations using equilibrium versus nonequilibrium methods. Analysis of heat flux autocorrelation functions show that very long ($6\times10^5\omega_p^{-1}$) time series are needed to reduce the noise level to allow $\lesssim2\%$ accuracy. The new simulations provide the first accurate data for $\Gamma \lesssim 1$. 
% allowing the tests of thermal conductivity models in a regime where they are predictive. 
%We test calculations of thermal conductivity using generalized Coulomb logarithms from the theories of Lee-More, Landau-Spitzer, Tanaka-Ichimaru, and Baalrud-Daligualt, and find that only the latter two can reproduce the trend of the MD data for $\Gamma<7$. 
This enables a test of the traditional Landau-Spitzer theory, which is found to agree with the simulations for $\Gamma \lesssim 0.3$. 
It also enables tests of theories to address moderate and strong Coulomb coupling. 
%The results indicate that transport theories must include the effect of particle correlations to properly model this range of coupling. 
Two are found to provide accurate extensions to the moderate coupling regime of $\Gamma \lesssim 10$, but none are accurate in the $\Gamma \gtrsim 10$ regime where potential energy transport and coupling between mass flow and stress dominate thermal conduction.

\end{abstract}

\maketitle

\section{Introduction}
A commonly used reference system in the study of strongly coupled plasmas is the one component plasma (OCP) model~\cite{1980PhR....59....1B}. The OCP consists of particles with charge $e$ and mass $m$ interacting via the Coulomb potential $e^2/r$ in a uniform, non-interacting background enforcing that the total system charge is zero. 
%The particles in the system are assumed to interact via the Coulomb potential $e^2/r$. 
Properties of the OCP can be completely characterized by the Coulomb coupling parameter, $\Gamma=e^2/ak_{\textrm{B}}T$, which is the ratio of the Landau length, $r_{\textrm{L}}=e^2/k_{\textrm{B}}T$, to the ion sphere radius, $a=(3/4\pi n)^{1/3}$. Here, $n$ is the number density and $T$ is the temperature. When $\Gamma \ll 1$ the plasma is weakly coupled and transport properties can be calculated by using traditional methods\cite{1970mtnu.book.....C,ferziger1972mathematical}.
When $\Gamma \gtrsim 1$ the plasma is strongly coupled and transport properties are influenced by complicated many-body correlations that are difficult for analytic theory to describe\cite{2006PhRvL..96f5003D}.

In practice, strong coupling directly influences the value of transport coefficients such as diffusion\cite{PhysRevA.11.1025,2006PhRvL..96f5003D}, shear viscosity\cite{PhysRevA.12.1106,2014PhRvE..90c3105D}, and thermal conductivity\cite{1978PhRvA..18.2345B,1998PhRvL..81.1622D,2004PhRvE..69a6405D}. 
The lack of accurate transport coefficients often presents a key uncertainty when modeling plasmas that can reach strong coupling regimes, such as hydrodynamic simulations of dense plasmas encountered in the implosion of inertial confinement fusion (ICF) targets. For example, differences in the value of these coefficients can alter the way that shocks propagate in the compressed shell and fuel~\cite{2017PhPl...24d2702V}, and can change the heat transport during the acceleration phase of the implosion (via direct-drive laser ablation) and heating of the compressing shell during the formation of a hot spot in the deuterium-tritium fuel~\cite{2019NucFu..59c2011H}. Accurate values of transport coefficients are critical for the design of such experiments~\cite{2018PhPl...25e6306H,2014PhRvE..89d3105H}. 
Transport coefficients at strong coupling are also important in more basic physics scenarios where our knowledge of transport informs our understanding of the cooling of neutron stars~\cite{10.1111/j.1365-2966.2007.12301.x}, heat transfer in dusty plasmas~\cite{PhysRevLett.95.025003}, and the dynamics of ultra cold neutral plasmas~\cite{PhysRevLett.109.185008,PhysRevX.6.021021}.  

Transport properties of the strongly coupled OCP have previously been studied theoretically~\cite{1978PhRvA..18.2345B,1985PhRvA..32.2981M,1986PhRvA..34.4163T,2013PhRvL.110w5001B} and using molecular dynamics (MD) simulations~\cite{1978PhRvA..18.2345B,1998PhRvL..81.1622D,2006PhRvL..96f5003D,2004PhRvE..69a6405D,2014PhRvE..90c3105D}. 
Comparing these predictions is valuable because the OCP model isolates the influence of strong coupling in a simplified model. 
Strongly coupled plasmas found in nature and the laboratory are often influenced by other physical effects that present their own complications for theory, such as mixtures of multiple species and the degeneracy of electrons in dense plasmas. 
%of without the additional complications encountered in mixtures or systems with electron degeneracy. 
Theories for the more complex systems, such as dense plasmas, often employ assumptions regarding strong Coulomb coupling that can be tested using the OCP~\cite{2016PhRvL.116g5002D}. 
%Therefore, tests of the OCP can influence the calculation of transport coefficients for physical materials. 
In this way, tests using the OCP model contribute to advancing the description of physical materials. 
%Theoretical models of the thermal conductivity have yet to be compared against MD simulations of the OCP, 
This work provides such a comparison for thermal conductivity, though it should also be noted that comparisons between multicomponent models and quantum molecular dynamics simulations have also been made for specific materials~\cite{2019NucFu..59c2011H}.

\begin{figure}
\includegraphics[width=\columnwidth]{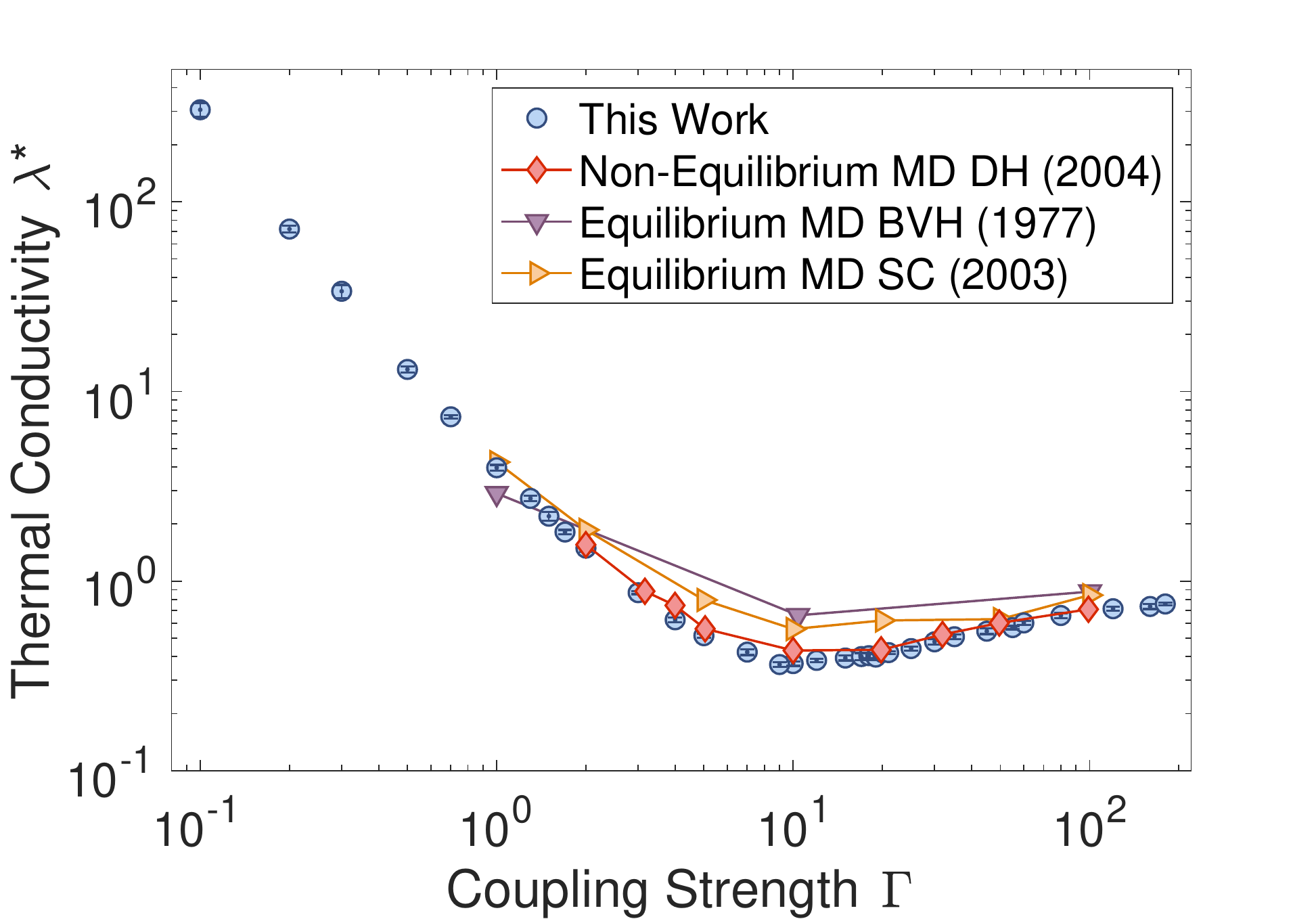}
\caption{\label{lambdas} Thermal conductivity ($\lambda^*=\lambda/n\omega_p k_B a^2$) from this work compared with those calculated by Donko and Hartmann (DH)\cite{2004PhRvE..69a6405D}, Bernu, Vieillefosse, and Hansen (BVH) \cite{1977PhLA...63..301B} and Salin and Caillol (SC) \cite{2003PhPl...10.1220S}. DH utilized a non-equilibrium method, while the remaining three used equilibrium MD.}
\end{figure}

%Thermal conductivity of the OCP has previously been studied using MD simulations.~\textcolor{red}{[citation]}
Our results advance the previous simulation efforts by extending the data set to both weaker and stronger coupling, and by improving the accuracy. 
% there are issues which limit the ability of these simulations to constrain models. For example, 
Previous MD simulations have used both equilibrium~\cite{1977PhLA...63..301B,2003PhPl...10.1220S} and non-equilibrium~\cite{2004PhRvE..69a6405D,1998PhRvL..81.1622D} methods. 
The published results differ by up to 30\% for $\Gamma\lesssim10$, with data from non-equilibrium MD considered to be the most accurate to-date~\cite{2004PhRvE..69a6405D}; see Fig.~\ref{lambdas}. Additionally, due to limitations of the  non-equilibrium MD method, no reliable data for thermal conductivity exists for $\Gamma\lesssim1$ and are limited by the number of particles required in the simulation, possibly needing $10^6$ to access low values of $\Gamma$\cite{doi:10.1063/1.873824,2004PhRvE..69a6405D,1998PhRvL..81.1622D}. 
Although results of the non-equilibrium method are considered the most accurate to-date, results of the equilibrium method are in many ways preferable because they provide access to the separate kinetic, potential, and virial contributions to the thermal conductivity. 
The equilibrium method is based on the Green-Kubo relation\cite{evans2007statistical}
\begin{equation}\label{cumint}
\lambda=\lim_{\tau\to\infty}\frac{1}{V k_B T^2}\int_0^\tau \langle \vc{j}(0)\cdot\vc{j}(t)\rangle dt,
\end{equation}
where
\begin{eqnarray}\label{jeq}
\vc{j}=\sum_{i=1}^N\bigg(\underbrace{\vc{v}_i\frac{1}{2}m|\vc{v}_i|^2}_{\textrm{kinetic}}+\underbrace{\frac{1}{2}\sum_{j\ne i}^N \phi(r_{ij})\vc{v}_i}_{\textrm{potential}}\nonumber \\
+\underbrace{\frac{1}{2}\sum_{j\ne i}^N (\vc{r}_{ij}\cdot\vc{v}_i)\vc{F}_{ij}}_{\textrm{virial}}\bigg)
\end{eqnarray}
is the heat flux. Here, $\vc{v}_i$ and $m$ are the velocity and mass of particle i, $\vc{F}_{ij}$ and $\vc{r}_{ij}$ are the force and vector between particles i and j, $\phi$ is the electrostatic potential, and $\langle \vc{j}(0)\cdot\vc{j}(t)\rangle$ is the heat flux autocorrelation function (hereafter ACF), where the angle brackets denote an ensemble average at equilibrium\cite{evans2007statistical}.
The values of thermal conductivity reported in this paper are given in terms of the dimensionless quantity $\lambda^*=\lambda/n\omega_p k_B a^2$. 

From Eq.~(\ref{jeq}), the components of the ACF can be calculated separately. The first of these is the kinetic energy flux which dominates at weak coupling and is the sole component considered in traditional weakly coupled plasma transport calculations. 
The ``potential'' term is associated with the potential energy transported by the mass flow and the ``virial'' term is associated with the coupling of stress with mass flow. 
% arise from the potential energy transported by the mass flow and the coupling of the stress with the mass flow, respectively, and both 
Both of these terms are expected to dominate at strong coupling, as they do in liquids~\cite{Eu}. 
This decomposition allows the relative importance of each term to be determined as a function of $\Gamma$, for the first time, in the OCP.
Similar decompositions have been recently applied to equilibrium MD simulations of the screened Yukawa OCP, both with and without a magnetic field~\cite{2015PhRvE..92f3105O}. 
Related MD simulation work has also been carried out on the 2D Yukawa system in the context of dusty plasmas\cite{PhysRevE.79.026401,PhysRevE.85.046405}. 
Here, we focus on the 3D classical OCP. 

In this paper, we identify issues that lead to uncertainty in equilibrium MD simulations of thermal conductivity, and provide a new dataset over a broad range of coupling strength $0.1\le\Gamma\le180$. 
The results are used to test the Landau-Spitzer model~\cite{1953PhRv...89..977S}, as well as to test the treatment of strong coupling in the Effective Potential Theory (EPT)~\cite{2013PhRvL.110w5001B} and theories of Tanaka-Ichimaru~\cite{1986PhRvA..34.4163T} and Lee-More~\cite{1984PhFl...27.1273L}. This comparison provides the first test of the widely used Landau-Spitzer (Spitzer-H{\"a}rm) thermal conductivity against MD data. This comparison allows direct comparison with theory over a larger range of $\Gamma$ than what has been possible experimentally\cite{PhysRev.112.1,1975PhFl...18.1467G,PhysRev.157.138,2017NatSR...7.7015M}. 
Comparisons against these models places constraints on their domain of validity and illustrates the inability of any of these models to capture the behavior of the thermal conductivity coefficient when $\Gamma \gtrsim 10$.

This paper is organized as follows: The details of the equilibrium MD setup are given in Sec. IIA and a discussion of sources of error in calculations of thermal conductivity for the OCP are given in Sec. IIB. Section IIC presents the new MD data and discusses qualitative features of the ACF and conductivity components encountered at different values of $\Gamma$. Section III compares the MD results to values of thermal conductivity calculated using generalized Coulomb logarithms from each of the different theories. Section IV concludes the paper with a summary of the results.

\section{Equilibrium Molecular Dynamics Simulations\label{secmd}}

\subsection{Simulation setup}

\begin{table}
\caption{MD simulation parameters}
\begin{ruledtabular}
\begin{tabular}{ccc}
 &($\Gamma\ge0.5$)&($\Gamma<0.5$)\\
 \hline
Number of Particles& 5000& 20000 \\
Time Step& $0.01\omega_p^{-1}$ & $0.001\omega_p^{-1}$ \\
Simulation Length&$6\times10^5\omega_p^{-1}$& $1\times10^5\omega_p^{-1}$ \\[1ex] 
 \end{tabular}
\end{ruledtabular}
\end{table}

Equilibrium MD simulations were carried out using the code LAMMPS~\cite{1995JCoPh.117....1P}. Two simulations were run for each of the selected values of $\Gamma$ in the range $0.1\le\Gamma\le180$. Initialization at a chosen value of $\Gamma$ involved fixing the number of particles, which scales the size of the periodic cubic domain, followed by a 900-1000~$\omega_p^{-1}$ equilibration phase via a Nose-Hoover thermostat~\cite{1984JChPh..81..511N} to achieve the desired temperature. 
Here, $\omega_p^{-1} = (m/4\pi e^2 n)^{1/2}$ is the angular plasma period, which sets the dimensionless timescale for the simulations. 
The particles interacted via the Coulomb potential, which was solved using the particle-particle-particle-mesh method on a $50\times50\times50$ mesh with a short range force cutoff of $5a$. Table~I gives the values of the  particle number, time step, and simulation run time used. 
%for $\Gamma\ge0.5$ and $\Gamma<0.5$. 
For low $\Gamma$, a smaller time step and larger number of particles were used to ensure accurate energy conservation and collisionality. 
However, due to the greater computational difficulty and less of a need to capture long-time oscillations of the ACF for $t \gtrsim 30\omega_p^{-1}$ (see discussion in Sec. IIC), the total number of plasma periods simulated was shorter at low $\Gamma$.  
The heat flux time series defined in Eq.~(\ref{jeq}), was calculated by summing over all particles in the simulation at a frequency of once every $0.1\omega_{p}^{-1}$.

\subsection{Sources of error}

\begin{figure}
\includegraphics[width=\columnwidth]{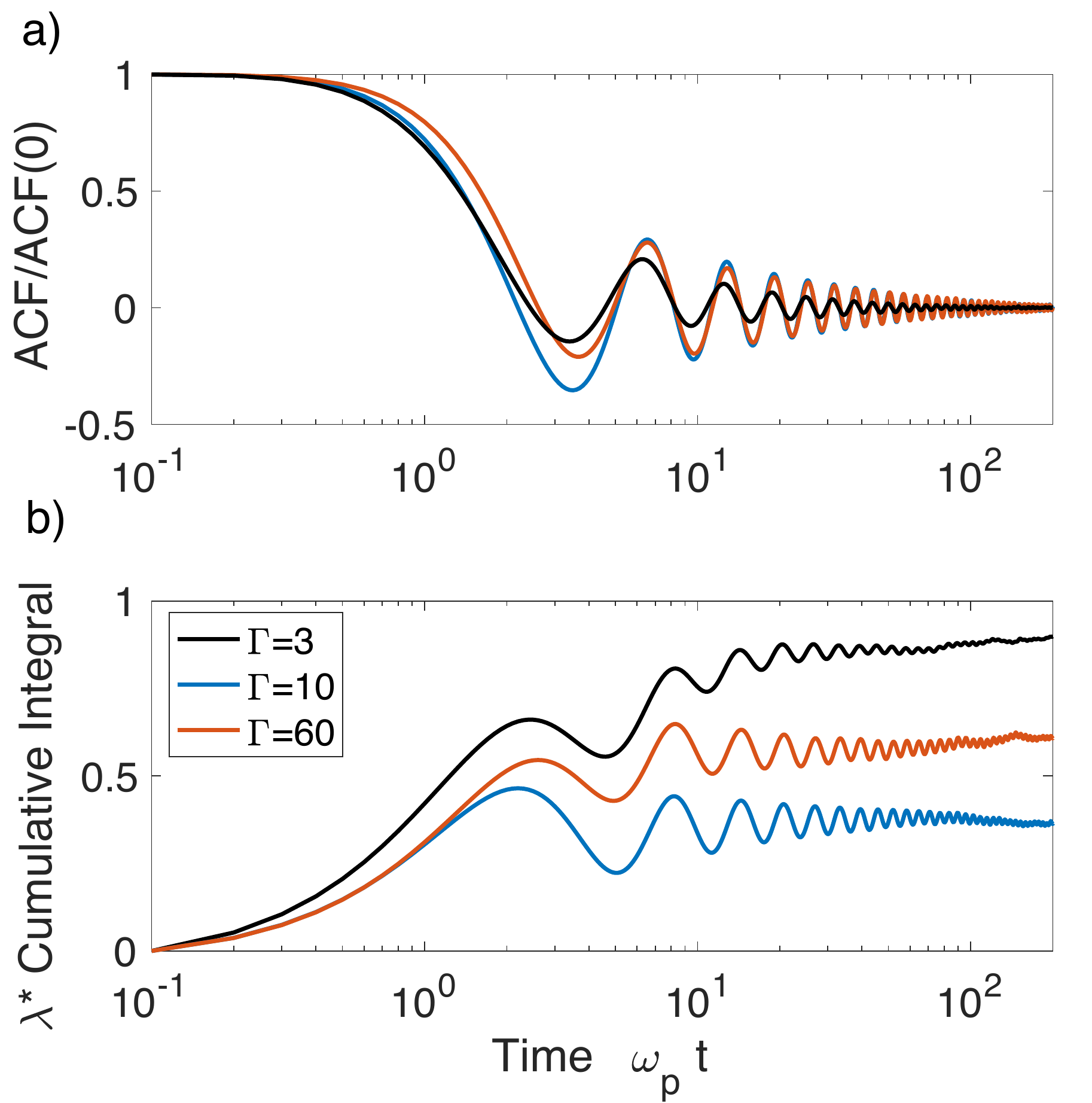}
\caption{\label{figACF}a) ACFs for $\Gamma=$ 3, 10, and 60. b) The corresponding cumulative integral from Eq.~(\ref{cumint}) as a function of integration time $\tau$.  }
\end{figure}

For $\Gamma \gtrsim 1$, the ACF oscillates for 100's of plasma periods as it slowly decays in amplitude, as shown in Fig.~\ref{figACF}a. 
% The result of the long-time oscillatory behavior of the ACF is that the 
Correspondingly, the cumulative integral, shown in Fig.~\ref{figACF}b, also oscillates as it asymptotes to a final value when $t\gtrsim100\omega_p^{-1}$. 
These slowly decaying oscillations have been predicted to have an envelope amplitude that scales as $1/\sqrt{t}$~\cite{1985PhRvA..32.2981M}. 
Due to the importance of the ACF at $t\gtrsim100\omega_{p}^{-1}$, the cumulative integral can encounter noise resulting from the finite length of the time series before convergence is obtained. 
This long time behavior in the ACF makes the thermal conductivity of the OCP more computationally demanding to calculate than other transport coefficients, such as diffusion\cite{2006PhRvL..96f5003D} or shear viscosity\cite{2014PhRvE..90c3105D}.
It also is more demanding than computing the thermal conductivity in systems that interact through shorter-range potentials, such as the $\kappa=2$ Yukawa OCP\cite{2015PhRvE..92f3105O}, in which the ACF decays more rapidly. 
In each of these cases, the corresponding ACF decays to near zero within 10's of $\omega_p^{-1}$~\cite{2014PhRvE..90c3105D,2015PhRvE..92f3105O}, minimizing the required integration over time intervals where the noise level is relatively large compared to the signal.

The primary difficulty encountered when computing the ACF is that only a finite time series for $\vc{j}(t)$ can be computed in the simulation. The equilibrium ensemble average ($\langle \vc{j}(0)\cdot\vc{j}(t)\rangle$ in Eq.~(\ref{cumint})) is replaced with an average of finite time and spatial extent,
\begin{equation}
\langle \vc{j}(0)\cdot\vc{j}(t)\rangle\approx C_\tau(t)\equiv\frac{1}{\tau}\int_0^\tau ds \ \vc{j}(s)\cdot\vc{j}(s+t),
\end{equation}
where $t$ is the time lag in the ACF integral and $\tau$ is the length of the time series in the integrand. Typically, the time series is padded with zeros on each end to facilitate the calculation of the time lag $t$ while still integrating $s$ over the interval $[0,\tau]$. This means that the ACF at time lag $t$ is calculated using a time series with effective length $\tau-t$ and has an error associated with that effective length. For ACF time lags of interest in this work $t\ll\tau$.

Estimates of the error induced in the ACF by approximating an infinite time series by a finite one were first provided by Zwanzig and Ailawadi~\cite{1969PhRv..182..280Z}.
Assuming a Gaussian processes, they estimated that the second moment of the deviation from the exact value of the correlation is 
\begin{equation}\label{eqvar}
\la \Delta(t_1)\Delta(t_2) \ra \approx \frac{2\tau_e}{\tau}[C_\infty(0)]^2,
\end{equation}
where 
\begin{equation}
\Delta(t) \equiv C_{\tau}(t)-C_\infty,
\end{equation}
$C_\infty$ is the exact correlation function for an infinite time series, $\la ...\ra$ is once again the ensemble average, and $\tau_e$ is an estimate of the $1/e$ decay time of the correlation function.

\begin{figure}
\includegraphics[width=\columnwidth]{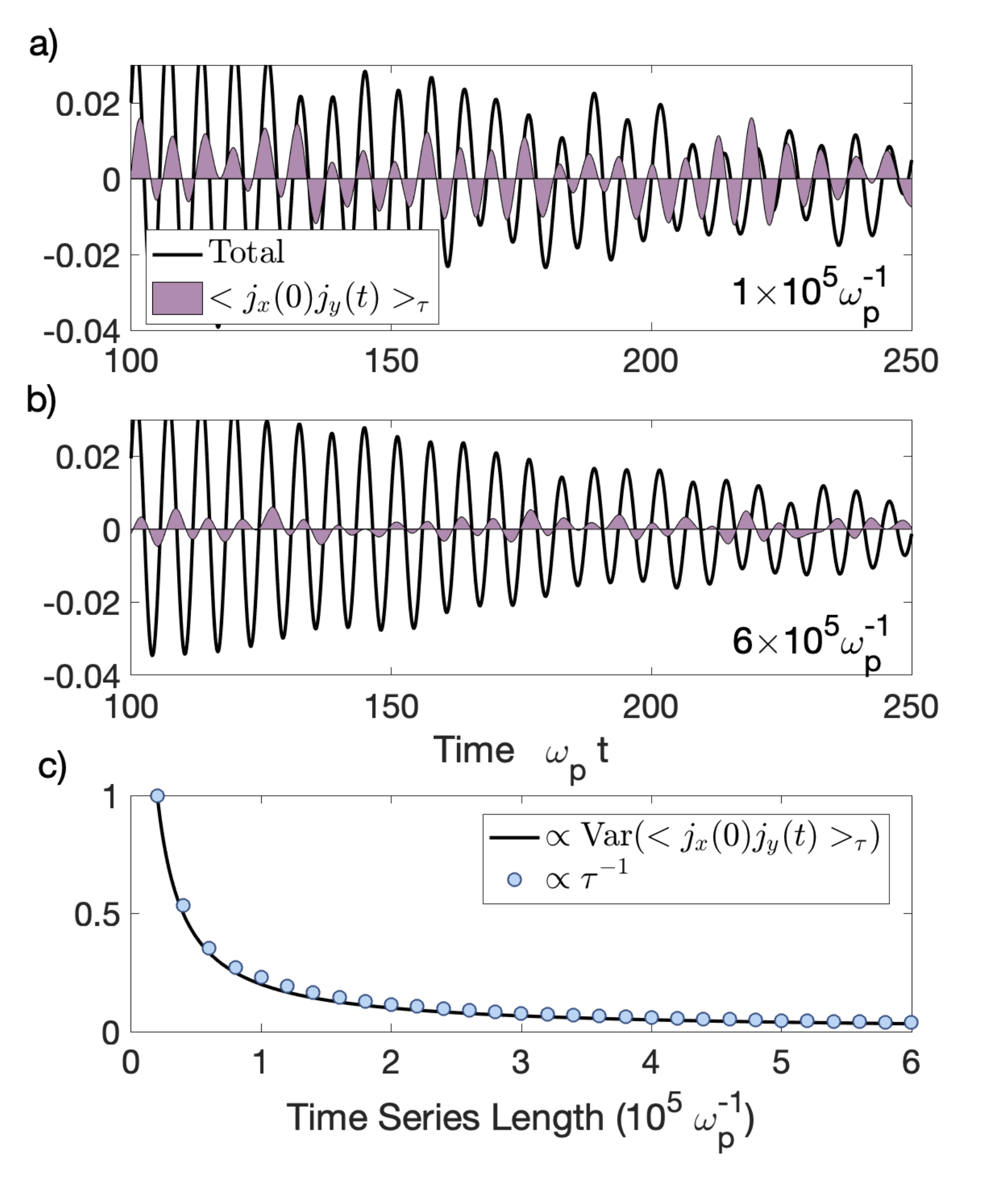}
\caption{\label{figNoise} (a) The total ACF compared to the cross correlation $\la j_x(0)j_y(t) \ra_\tau$, used as an estimate for the noise level, for a $1\times10^5\omega_p^{-1}$ time series. (b) The same as (a), but with a $6\times10^5\omega_p^{-1}$ time series. (c) The variance of the cross correlation compared with the scaling of the error predicted by Eq.~(\ref{eqvar}).  }
\end{figure}

The noise level in the ACF computed from the simulations can be estimated for comparison with Eq.~(\ref{eqvar}) by taking the cross correlation of different vector components of $\vc{j}(t)$. 
Since the cross correlation should be zero in the thermodynamic limit, its value for a finite time series is a good estimator of the noise level~\cite{2012JChPh.137v4111H}. 
Figures~\ref{figNoise}a and b show the value of the total ACF (for $\Gamma=30$) alongside the value of the cross correlation $\langle j_x(0)j_y(t)\rangle_\tau=\frac{1}{\tau}\int_0^\tau j_x(s)j_y(s+t)ds $ for time series of length $\tau = 1\times10^5\omega_p^{-1}$ and $6\times10^5\omega_p^{-1}$, respectively. For $1\times10^5\omega_p^{-1}$, the level of fluctuations of the cross correlation approaches the total value of the ACF itself by $t=200\omega_p^{-1}$, however, for $6\times10^5\omega_p^{-1}$ the noise is significantly reduced. This level of improvement is in agreement with the analysis of Zwanzig and Ailawadi. Figure~\ref{figNoise}c shows the $1/\tau$ scaling of Eq.~(\ref{eqvar}) plotted against the scaling of the variance of the cross correlation calculated using several time series of varying length. The figure shows that the decrease in noise with time series length is in agreement with the prediction for $\Gamma=30$. 
This agreement is characteristic of that for the entire range of $\Gamma$ studied. 
This trend suggests that doubling the simulation length will result in a decrease of noise by a factor of two. 

Previous equilibrium MD simulations by Bernu and Vieillefosse~\cite{1978PhRvA..18.2345B} and Salin and Caillol~\cite{2003PhPl...10.1220S} ran for $\sim400\omega_p^{-1}$ and $\sim8000\omega_p^{-1}$, respectively. 
The estimates above indicate that significant errors associated with noise in the cumulative integral are likely with this short of a time series. 
This may be responsible for the observed discrepancies with non-equilibrium MD results; see Fig.~\ref{lambdas}. 
%With such short time series, integration of oscillations in the autocorrelation tail would have been inaccurate leading to the observed discrepancies with non-equilibrium MD data.

\subsection{Simulation results}

\begin{figure}
\includegraphics[width=\columnwidth]{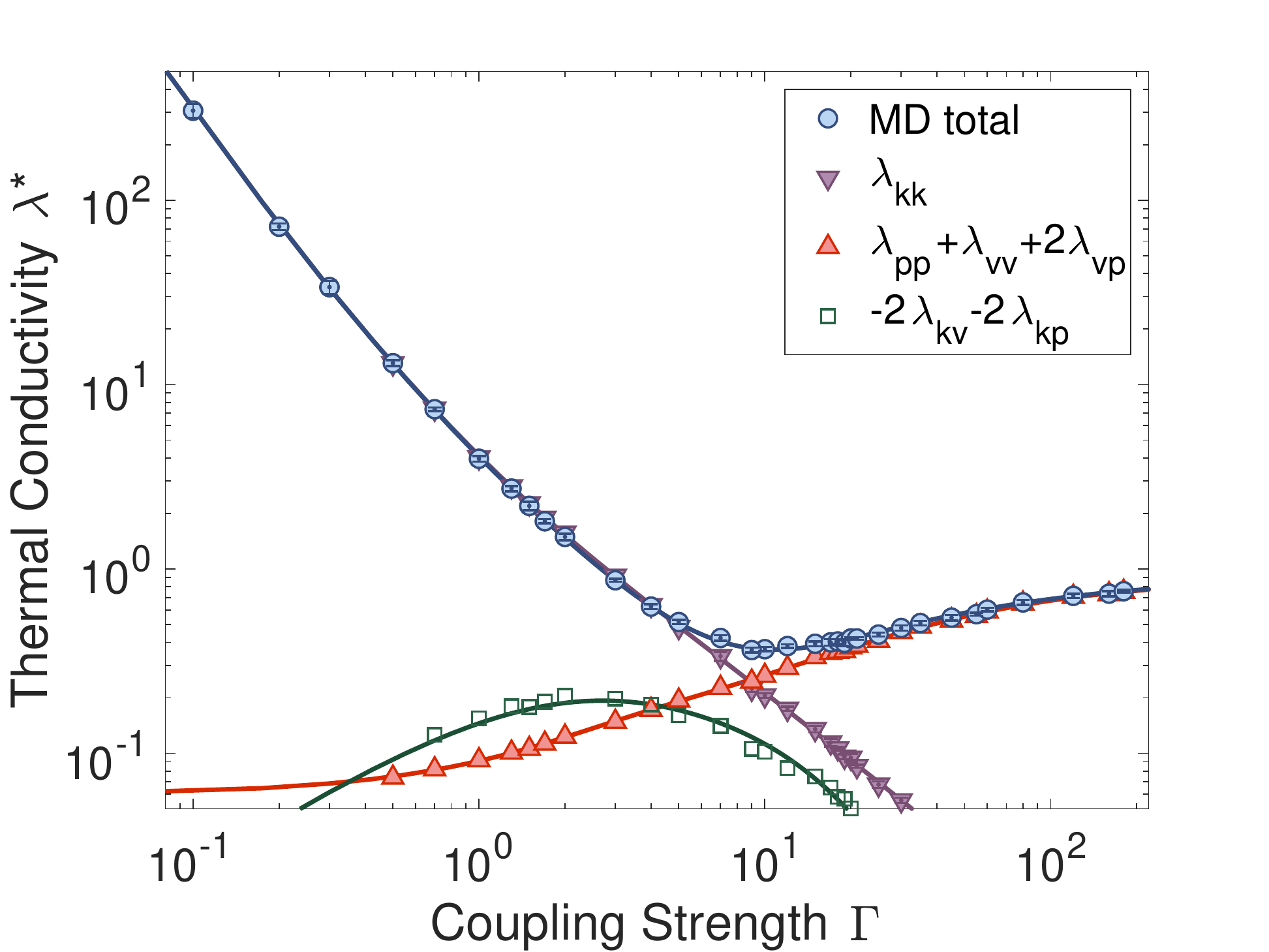}
\caption{\label{figLambda}Equilibrium MD data for the total thermal conductivity and its kinetic ($\lambda_{\textrm{kk}}$), potential and virial ($\lambda_{\textrm{pp}}+\lambda_{\textrm{vv}}+2\lambda_{\textrm{vp}}$), and kinetic cross components ($-2\lambda_{\textrm{kp}}-2\lambda_{\textrm{kv}}$). Error bars are shown for the total and kinetic parts and are approximately the size of the marker. The solid lines represent fits to the separate components described in the text. }
\end{figure}

The value of the thermal conductivity was calculated from the simulations described in Sec.~2A taking consideration of the sources of error described in Sec.~2B. 
For each value of $\Gamma$, the time interval of the cumulative integral that was taken as the asymptotic value was selected to avoid contamination by noise. 
For the case of $\Gamma\ge0.5$, the cumulative integral of each time series was calculated out to $t=150\omega_p^{-1}$. 
At this time, oscillations in the cumulative integral were observed to have decreased significantly and the cumulative integral of the noise had yet to significantly alter the value. 
However, since oscillations are still present at the level of $\sim5\%$ at $t=100\omega_p^{-1}$ (see Fig.~\ref{figACF}B), an averaging window of the cumulative integral over $[100\omega_p^{-1},150\omega_p^{-1}]$ was used to calculate the final value for each time series. 
The reported values did not change significantly when using alternate averaging windows of $[100\omega_p^{-1},200\omega_p^{-1}]$ or $[150\omega_p^{-1},200\omega_p^{-1}]$. 
To further improve statistics, data from all three spatial directions and both simulation runs were used to independently calculate the value of the thermal conductivity, resulting in six independent time series for each value of $\Gamma$. These were averaged to determine the value and standard deviation reported in Fig.~\ref{figLambda} and Table II.

\begin{figure}
\includegraphics[width=\columnwidth]{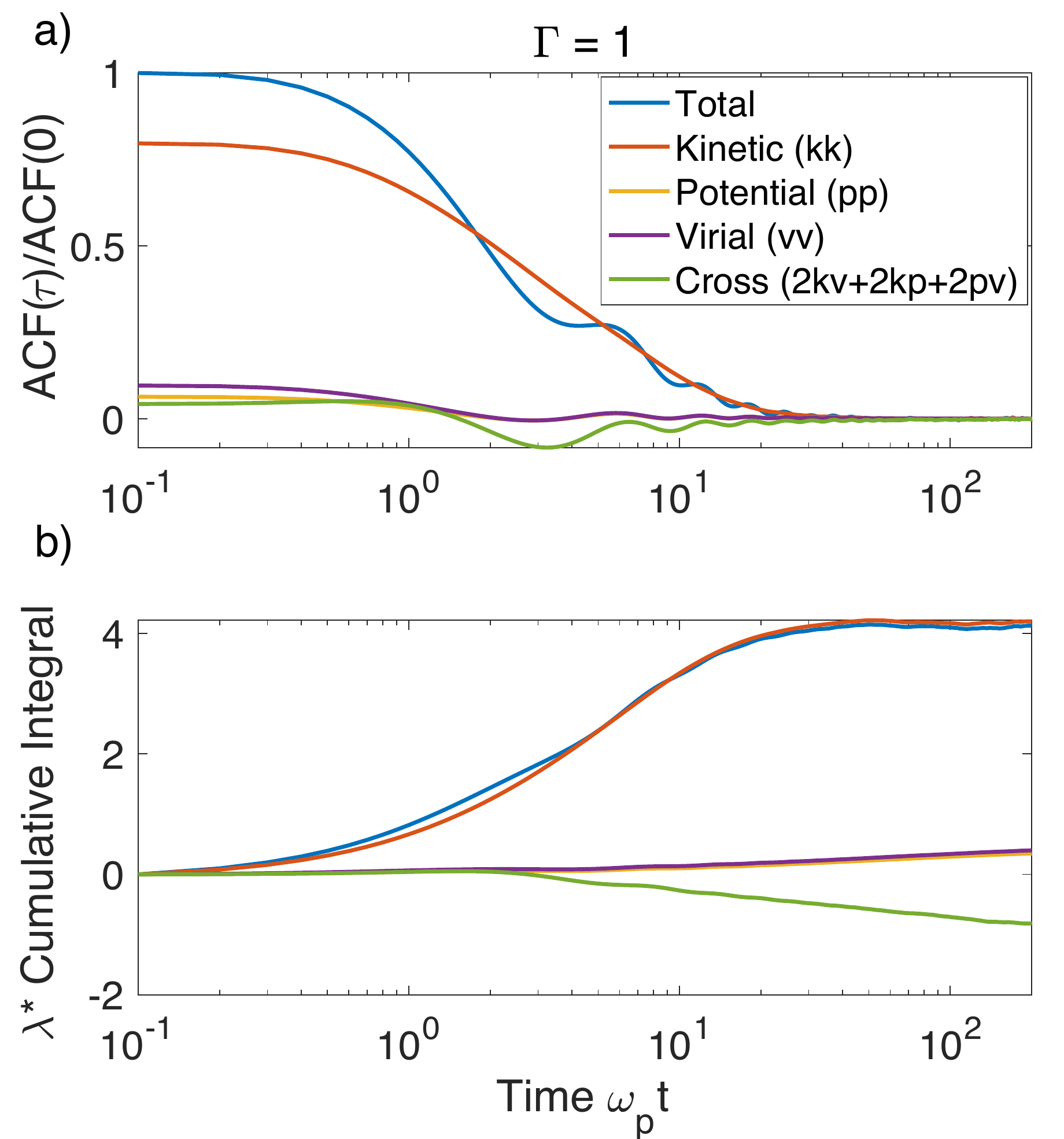}
\caption{\label{figParts1} (a) The total ACF and its components. (b) The corresponding cumulative integral of the total ACF and its components.  }
\end{figure}

\begin{figure}
\includegraphics[width=\columnwidth]{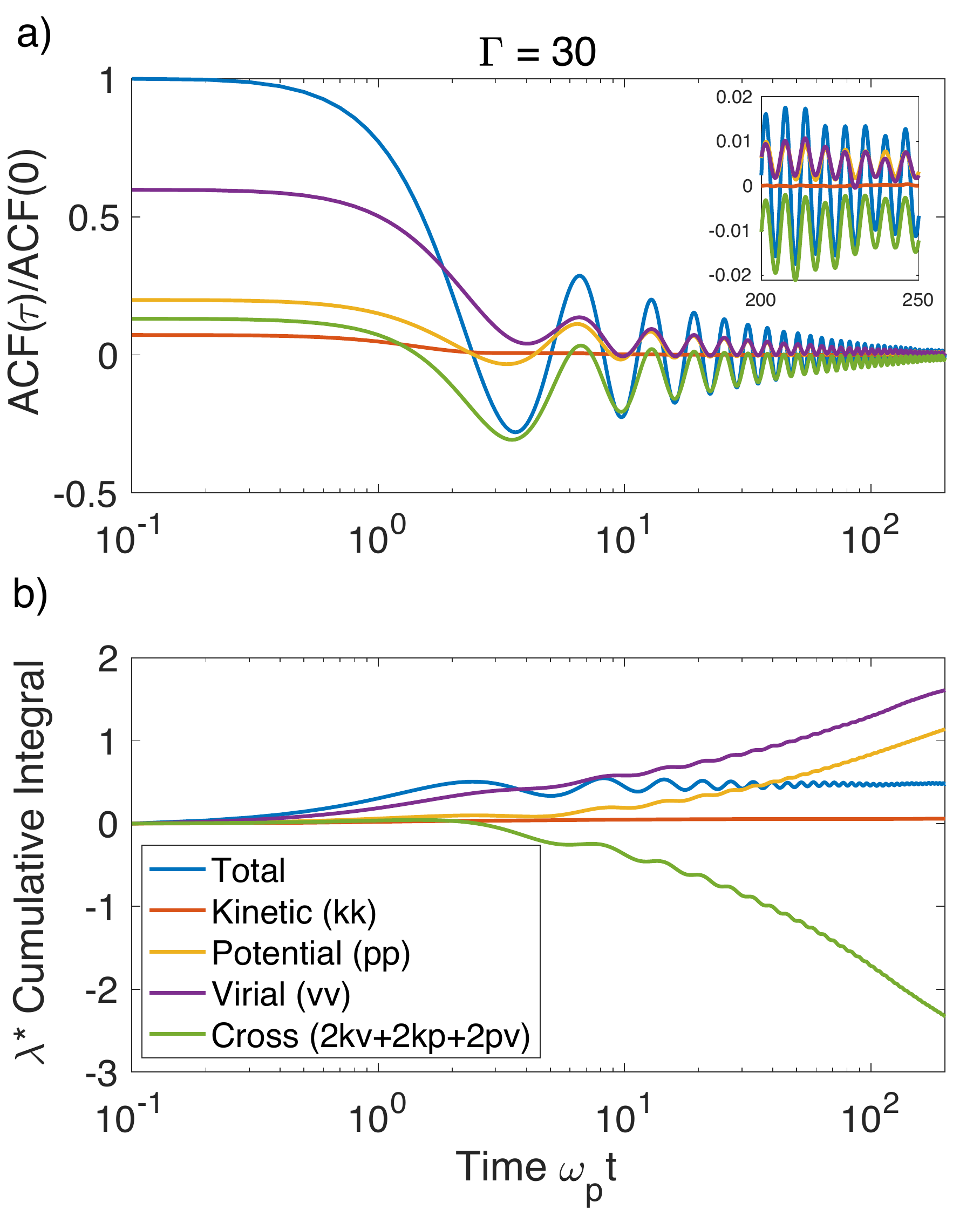}
\caption{\label{figParts} (a) The total ACF and its components. (b) The corresponding cumulative integral of the total ACF and its components. }
\end{figure}

In general, long time oscillations decrease with decreasing $\Gamma$; for example, compare Fig.~\ref{figParts1}a and Fig~\ref{figParts}a. 
%\textcolor{blue}{How do these figures which show data for $\Gamma = 1$ and $30$ help you make a point that you said is valid when $\Gamma < 0.5$?}
Due to this observation, a different method was used to process data for $\Gamma<0.5$. In these cases the cumulative integral saturated around $100-200\omega_p^{-1}$ providing the thermal conductivity coefficient for each of the six time series discussed in the previous paragraph. For these values of $\Gamma$, only the value of thermal conductivity at saturation is used, omitting the averaging window of the cumulative integral. The six values of thermal conductivity were then averaged and are also reported in Fig.~\ref{figLambda} and Table II with their standard deviation. 
%\textcolor{blue}{I don't understand how this is different? Is this where you mean to say that a shorter simulation time was used?}

\begin{table}
\caption{Values of thermal conductivity calculated using 6 independent heat flux time series along with the standard deviation shown to two significant digits.}
\begin{ruledtabular}
\begin{tabular}{cccccc}
$\Gamma$ & $\lambda^*$ & $\Gamma$ & $\lambda^*$ & $\Gamma$ & $\lambda^*$ \\ [0.5ex] 
 \hline
 0.1&306$\pm$26&4.0&0.625$\pm$0.019&21.0&0.4197$\pm$0.0073\\
 0.2 &71.9$\pm$3.0&5.0&0.515$\pm$0.014&25.0&0.441$\pm$0.011\\ 
 0.3 &33.8$\pm$2.7&7.0&0.422$\pm$0.014&30.0&0.478$\pm$0.016\\ 
 0.5 &13.07$\pm$0.48&9.0&0.363$\pm$0.011&35.0&0.509$\pm$0.015\\ 
 0.7 &7.35$\pm$0.16&10.0&0.3672$\pm$0.0096&45.0&0.544$\pm$0.019\\ 
 1.0 &3.96$\pm$0.14&12.0&0.3818$\pm$0.0083&55.0&0.569$\pm$0.011\\ 
 1.3 &2.728$\pm$0.089&15.0&0.393$\pm$0.012&60.0&0.602$\pm$0.014\\ 
 1.5 &2.20$\pm$0.12&17.0&0.400$\pm$0.017&80.0&0.658$\pm$0.020\\ 
 1.7 &1.815$\pm$0.049&18.0&0.404$\pm$0.013&120.0&0.715$\pm$0.018\\ 
 2.0 &1.492$\pm$0.059&19.0&0.398$\pm$0.015&160.0&0.735$\pm$0.024\\
 3.0&0.869$\pm$0.016&20.0&0.4196$\pm$0.0053&180.0&0.757$\pm$0.013\\ [1ex] 
 \end{tabular}
\end{ruledtabular}
\end{table}

In addition to the total thermal conductivity, the Green-Kubo formalism allows the analysis of different contributions to the ACF and their cumulative integrals. The thermal conductivity can be split into components due to products of like and unlike parts of the heat flux in the autocorrelation of Eq.~(\ref{cumint})
% and can be written in the form 
\begin{equation}
\lambda=\lambda_{\textrm{kk}}+\lambda_{\textrm{pp}}+\lambda_{\textrm{vv}}+2\lambda_{\textrm{kp}}+2\lambda_{\textrm{kv}}+2\lambda_{\textrm{pv}}
\end{equation}
where the subscripts k (kinetic), p (potential), and v (virial) indicate the heat flux terms used in the autocorrelation (e.g. for $\lambda_{\textrm{kv}}$ the cross correlation $\langle \vc{j}_{\textrm{kinetic}}(0)\cdot\vc{j}_{\textrm{virial}}(t)\rangle$ is used)~\cite{2015PhRvE..92f3105O}. 
Figures~\ref{figParts1}a and \ref{figParts}a show the different components of the ACF and their cumulative integral for $\Gamma=1$ and 30, respectively. The kinetic (kk), potential (pp), and virial (vv) parts result from the products of like terms, while the cross term is the sum of all products of unlike terms, being primarily composed of the vp term. The values shown for $\Gamma=30$ are qualitatively similar to those of other values of $\Gamma\gtrsim20$ while in the cases where $\Gamma<1$ only the kinetic part is important. The kinetic part $\lambda_{\textrm{kk}}$, shown in Fig.~\ref{figLambda}, is the dominant term from low $\Gamma$ to about $\Gamma=5$. 
This is the regime where the kinetic energy flow determines the transport coefficients~\cite{ferziger1972mathematical}. 
As the coupling increases, the increased collisionality reduces the rate of mass flow and hence the value of $\lambda_{\textrm{kk}}$.

At $\Gamma\gtrsim10$ plasma transport becomes more liquid-like as strong correlations become dominant. 
Here, the pp, vv, and pv terms in the ACF are the largest in magnitude. 
Inspection of the $\Gamma=30$ ACF in Fig.~\ref{figParts}a indicates that the potential, virial, and cross terms oscillate in phase. The potential and virial terms oscillate above zero, and the cross term oscillates primarily below zero. 
%The sign of the respective terms indicates that an increase in the potential energy flux due to particle motion is correlated with a corresponding decrease in the virial term. 
%This suggests that an increase in heat flow due to potential energy flux decreases the heat flow due to coupling of the mass flow with the fluid stress (this has been described as the flow of stress related work~\cite{Eu}) and vice versa. 
The sign of the respective terms indicates that an increase in heat flow due to potential energy flux is correlated with a decrease in heat flow due to coupling of the mass flow with the fluid stress (this has been described as the flow of stress related work~\cite{Eu}) and vice versa, suggesting a possible interplay between these features. The persistence of oscillations for 100s of $\omega^{-1}_p$ suggests that long range collective behavior plays an important role in the oscillations.  

A comparison of Fig.~\ref{figParts}a with Fig.~2 of Ref.~\cite{doi:10.1002/ctpp.201500101} highlights the role of the long range potential in comparison with the shorter-range $\kappa=2$ Yukawa OCP (hereafter YOCP) at values of $\Gamma$ where liquid like behavior sets in ($\Gamma=30-300$ in Fig.~2 of Ref.~\cite{doi:10.1002/ctpp.201500101}). The most notable difference is the absence of persistent oscillations in the pp, vv, and pv terms (labeled pp, cc, and pc in Ref.~\cite{doi:10.1002/ctpp.201500101}) of the YOCP. Compared to the YOCP, the magnitude of the pp and vv components of the ACF are closer in magnitude, possibly owing to the larger contribution to the potential term from long range interactions in the OCP. Additionally, the pv term, which is the largest in magnitude in the OCP, is small in the YOCP. Furthermore, nearly all components are positive, meaning that no cancelation of terms takes place in the total value of the ACF. The suppression of oscillations in the total ACF with increasing $\kappa$ was previously observed in Ref.~\cite{2004PhRvE..69a6405D}. These observations lend additional evidence to the assertion that the observed behavior is associated with the long range nature of the bare Coulomb potential compared to the Yukawa OCP. 

%The long range Coulomb force allows the motion of single particles to contribute to the fluid stress to a greater extent than particles which interact via short range potentials. This behavior is not present in the $\kappa=2$ Yukawa OCP where the pv term is small compared to the vv term~\cite{doi:10.1002/ctpp.201500101}. This is in agreement with the assertion that the observed behavior is associated with the long range nature of the bare Coulomb potential compared to neutral fluids and the Yukawa OCP. 

The oscillation of the pp, vv, and pv terms either strictly above or below zero prevents the calculation of $\lambda_{\textrm{pp}}$, $\lambda_{\textrm{vv}}$, and $\lambda_{\textrm{vp}}$ in isolation since the magnitude of their cumulative integrals increase in time, as demonstrated in Fig.~\ref{figParts}b, and does not saturate before noise dominates the value of the ACF. 
This is a feature resulting from the long timescale needed for the complete decay of these oscillations.
However, the sum of these oscillating components $\lambda_{\textrm{pp}}+\lambda_{\textrm{vv}}+2\lambda_{\textrm{vp}}$ converges, reaching an asymptotic value by $150\omega_p^{-1}$. 
For this reason, only the combinations $\lambda_{\textrm{kk}}$, $\lambda_{\textrm{pp}}+\lambda_{\textrm{vv}}+2\lambda_{\textrm{vp}}$ and $-2\lambda_{\textrm{kv}} - 2 \lambda_{\textrm{kp}}$ were computed. 
Figure~\ref{figParts}b shows that at $\Gamma=30$ the second of these contributions dominates the total thermal conductivity, since the kinetic term is negligible. 
In Fig.~\ref{figLambda}, this component is shown to be the dominant contribution at very strong coupling. 
The dominance of the potential and virial parts of the heat flux is also known to occur in liquids and dense gases~\cite{Eu}. 

At $\Gamma\approx9$, the thermal conductivity exhibits a minimum. 
Around this value of $\Gamma$, $\lambda_{\textrm{kk}}$ becomes smaller than $\lambda_{\textrm{pp}}+\lambda_{\textrm{vv}}+2\lambda_{\textrm{vp}}$, which increases with increasing $\Gamma$. 
The kinetic contribution becomes reduced due to increased collisionality and the potential and virial components become more important as liquid-like behavior sets in.
Also around this value of $\Gamma$, the value of the kinetic cross term $2\lambda_{\textrm{kv}}+2\lambda_{\textrm{kp}}$ reaches its maximum neagitive value, decreasing the value of the overall transport coefficient and playing an important role in the location of the minimum. 
This term becomes unimportant for $\Gamma \gtrsim 11$ due to the near-zero kinetic contribution to the heat flux.

For ease of use we provide a fit for each of $\lambda_{\textrm{kk}}$, $\lambda_{\textrm{pp}}+\lambda_{\textrm{vv}}+2\lambda_{\textrm{vp}}$, and $2\lambda_{\textrm{kv}}+2\lambda_{\textrm{kp}}$. Following Ref.~\onlinecite{2014PhRvE..90c3105D}, the kinetic part is first fit to the form $C/[\Gamma^{5/2}\ln(1+D\Gamma^{-3/2})]$ so that it is proportional to $C/\ln(D\Gamma^{-3/2})$ at low $\Gamma$. This is favorable since it is a trivial modification of the Landau-Spitzer form of the Coulomb logarithm (cf. Eq.~\ref{LSXI}) which is expected to be valid at weak coupling. Once the values for C and D are obtained, the remainder is fit with a Pad\'{e} approximation, so that the total fit takes the form 
\begin{equation}\label{fit}
\lambda^{*\textrm{fit}}_{\textrm{kk}}=\frac{C}{\Gamma^{5/2}\ln(1+D\Gamma^{-3/2})}\frac{p_1+p_2\Gamma+p_3\Gamma^2+p_4\Gamma^3}{q_1+q_2\Gamma+q_3\Gamma^2+q_4\Gamma^3}.
\end{equation}
The MD data for $\lambda_{\textrm{pp,pv,vv}} \equiv \lambda_{\textrm{pp}}+\lambda_{\textrm{vv}}+2\lambda_{\textrm{vp}}$ and $\lambda_{\textrm{kv,kp}} \equiv 2\lambda_{\textrm{kv}}+2\lambda_{\textrm{kp}}$ were fit with the Pad\'{e} approximation 
\begin{equation}\label{fitpv}
\lambda^{*\textrm{fit}}_{\textrm{pp,pv,vv}}=\frac{p_1\Gamma^3 + p_2\Gamma^2 + p_3\Gamma+ p_4}{\Gamma^3 + q_1\Gamma^2 + q_2\Gamma + q_3}
\end{equation}
and biexponential function
\begin{equation}\label{fitk}
\lambda^{*\textrm{fit}}_{\textrm{kv,kp}}=-p_1\big[\exp(-p_2\Gamma)- \exp(-p_3\Gamma)\big],
\end{equation}
respectively. The fit coefficients are given in Table III and the individual fits are plotted with the MD data in Fig.~\ref{figLambda}. 
Each fit shows excellent agreement with the components of the thermal conductivity, with the fits passing through each of the markers of their respective components, with one exception for Eq.~(\ref{fitk}). 
The sum of the fits $\lambda^{*\textrm{fit}} = \lambda^{*\textrm{fit}}_{\textrm{kk}}+\lambda^{*\textrm{fit}}_{\textrm{pp,pv,vv}}+\lambda^{*\textrm{fit}}_{\textrm{kv,kp}}$ is also in good agreement with the total thermal conductivity, with no discernible deviation from the MD data. 
%\textcolor{blue}{Is there a quick way to describe the accuracy of the fit? Perhaps just say that ``the fit for the total thermal conductivity is within $x\%$ of the value computed from MD over the entire range of $\Gamma$ simulated''?}
Since the fit asymptotes to the Landau-Spitzer result for $\Gamma \ll 1$, it is expected to extrapolate to the weakly coupled limit. 
Thus, it provides a convenient characterization of the thermal conductivity of the OCP spanning from weak coupling up to solidification, which occurs near $\Gamma = 170$\cite{PhysRevA.21.2087}.

\begin{table}
\caption{Fit parameters}
\begin{ruledtabular}
\begin{tabular}{cccc}
Fit Parameter & $\lambda^{*\textrm{fit}}_{\textrm{kk}}$ & $\lambda^{*\textrm{fit}}_{\textrm{pp,pv,vv}}$ & $\lambda^{*\textrm{fit}}_{\textrm{kv,kp}}$  \\ [0.5ex] 
& (Eq.~\ref{fit})&(Eq.~\ref{fitpv})&(Eq.~\ref{fitk})\\ [0.5ex] 
 \hline
 C&3.766&-&-\\
 D&1.52&-&-\\
$p_1$ &-0.1401&0.8835&0.2678\\ 
$p_2$ &0.1734&0.9607&0.0868\\ 
$p_3$ &1.836&11.28&0.9836\\ 
$p_4$ &0.0388&28.86&-\\
$q_1$ &-0.1401&33.27&-\\ 
$q_2$ &0.1939&-58.93&-\\  
$q_3$&1.74&486.4&-\\
$q_4$&0.09111&-&-\\ [1ex] 
 \end{tabular}
\end{ruledtabular}
\end{table}

%\section{Models of Thermal Conductivity}
\section{Comparison with Theoretical Models}

In this section, the treatments of strong coupling in models of thermal conductivity are tested against the MD data presented in Sec.~IIC. While several models of thermal conductivity for dense plasmas exist in the literature, many of these are formulated for a two component electron-ion plasma and treat the electron degeneracy to some extent~\cite{1985PhRvA..32.1790I,1984PhFl...27.1273L}. 
Nevertheless, the treatment of strong coupling in these models can be tested against the OCP by considering the classical one-component limit. 
%evaluating the modifications to the Coulomb logarithm in each model. 

\begin{figure}
\includegraphics[width=\columnwidth]{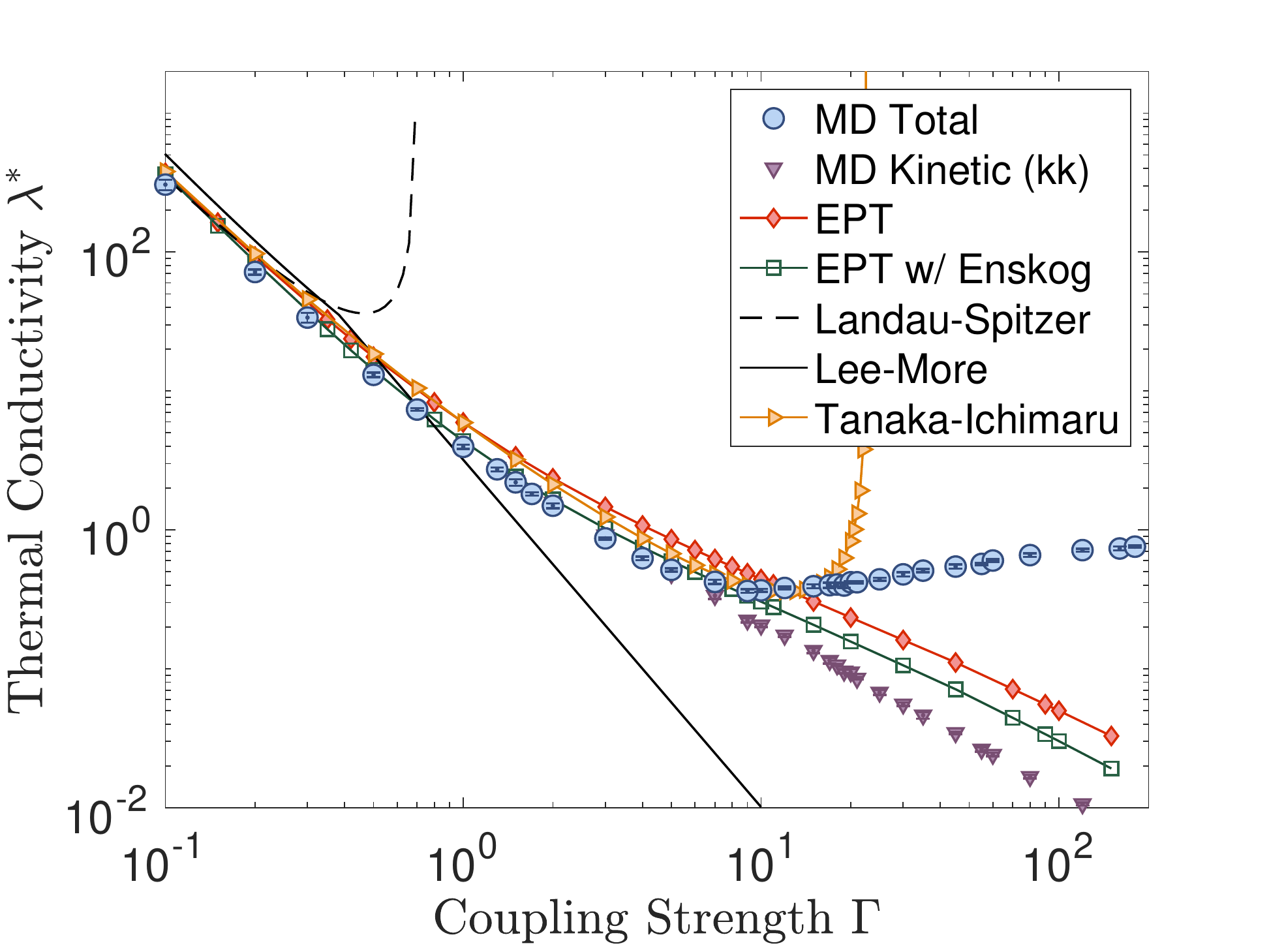}
\caption{\label{figTheory} A comparison of the calculated value of $\lambda^*$ using different generalized Coulomb logarithms to the total and kinetic parts of the thermal conductivity from MD simulations. }
\end{figure}

In each of the models considered, thermal conductivity can be cast in the form~\cite{ferziger1972mathematical}
\begin{equation}\label{lstar}
\lambda^*  = \frac{25}{4} \frac{\sqrt{\pi/3}}{\Gamma^{5/2} \Xi},
\end{equation}
where $\Xi$ is a generalized Coulomb logarithm. 
In this section, four models for the generalized Coulomb logarithm are considered: Landau-Spitzer~\cite{1953PhRv...89..977S}, Lee-More~\cite{1984PhFl...27.1273L}, Tanaka-Ichimaru~\cite{1986PhRvA..34.4163T}, and Baalrud-Daligualt (EPT)~\cite{2015PhRvE..91f3107B,2013PhRvL.110w5001B}. 
A summary of the model predictions is shown in Fig.~\ref{figTheory}. 

Of the models presented, none can reproduce the trend of the thermal conductivity for $\Gamma \gtrsim 10$ where the potential and virial parts of the thermal conductivity become dominant; compare Fig.~\ref{figTheory} and Fig.~\ref{figLambda}. 
This is expected because all of these models only consider transport associated with the kinetic energy (mass flow) of particles. 
%presented only consider binary interactions. In this treatment, only the kinetic part of the thermal conductivity due to the mass flow of particles is included. 
%Since they do not consider how the motion of the colliding test particle may change the interaction potential or fluid stress. 
Since they do not include the physics processes relevant to the potential or virial components, they should, at best, be able to reproduce the kinetic contribution in comparison to the MD simulations. 
%Agreement between MD and theory, at best, should only be expected for the kinetic part.

\subsection{Landau-Spitzer}

Landau-Spitzer is the textbook weakly coupled plasma theory\cite{spitzer1962physics}.  
%The traditional result can be obtained in the weakly coupled limit through the use of a 
It can be derived from a Boltzmann-type kinetic equation by modeling interactions as occurring via a Coulomb potential with a range cutoff at the Debye length, $\lambda_{\textrm{D}}$, when computing the momentum transfer cross section and expanding in the weakly coupled limit $\Gamma \ll 1$\cite{doi:10.1063/1.4875282}.  
The well-known result for a single species plasma is~\cite{1953PhRv...89..977S} 
\begin{equation}\label{LSXI}
\Xi_{\textrm{LS}} = 2\ln\Lambda=2\ln(\sqrt{3}\Gamma^{-3/2}),
\end{equation}
where $\Lambda$ is the plasma parameter. 
The thermal conductivity using Eq.~(\ref{LSXI}) in Eq.~(\ref{lstar}) is shown in Fig.~\ref{figTheory}. 
Good agreement with the MD data is observed at the weakest coupling parameters simulated ($\Gamma \lesssim 0.3$), while its value diverges as $\Gamma$ approaches approximately 0.7. 
This comparison represents the first test of the Landau-Spitzer thermal conductivity against MD data because it is the first time that MD simulations of thermal conductivity have been performed at weak enough coupling strength to reach this regime. 
It provides a high-accuracy test of the theory that would be difficult to achieve experimentally.

\subsection{Lee-More Non-Degenerate Limit}

The Lee-More model is one of the most commonly used models for electrical and thermal conductivity in radiation hydrodynamic simulations of high-energy density plasmas (see Ref.~\cite{2018NatSR...8.4525G} or Ref.~\cite{doi:10.1063/1.5097960} for recent examples). The model attempts to include the effects of electron degeneracy as well as the increased collisionality at strong coupling while reproducing the classical Spitzer conductivity scaling in the non-degenerate $\lambda_{\textrm{D}} > a$ limit~\cite{1984PhFl...27.1273L}. In the case where the Debye length is less than the interparticle spacing, the ion sphere radius is used in place of the Debye length. Additionally, the model limits the value of the Coulomb logarithm to be greater than 2. The form of the non-degenerate generalized Coulomb logarithm is
\begin{equation}\label{LMXI}
\Xi_{\textrm{LM}} =\textrm{max}\bigg\{\frac{1}{2}\ln\bigg[1+(\textrm{max}\{\sqrt{3}\Gamma^{-3/2},3/\sqrt{\Gamma}\})^2\bigg], 2\bigg\}.
\end{equation}

The thermal conductivity using Eq.~(\ref{LMXI}) in Eq.~(\ref{lstar}) is shown in Fig.~\ref{figTheory}. The value of $\lambda^*$ disagrees with the MD data for $\Gamma\gtrsim 1$, demonstrating that the replacement of the Debye length with the ion sphere radius is an insufficient model for accounting for the increased collision rate of strongly coupled plasmas. Realizing the under-prediction of thermal conductivity in simulations may play an important role in improving the agreement between high energy density experiments and simulations, particularly in simulations of solid density materials where strong coupling in electron-ion collisions is likely to occur.

\subsection{Tanaka-Ichimaru}

Tanaka and Ichimaru formulated a model for the thermal conductivity of a two component plasma composed of electrons and ions of charge $Ze$ based on linear response theory, which attempts to generalize the Lenard-Balescu equation to account for effects of electron degeneracy and strong Coulomb coupling~\cite{1985PhRvA..32.1790I}. Although their thermal conductivity model is formulated for a two component plasma, their treatment of strong coupling can be tested via an OCP Coulomb logarithm formulated in later work on shear viscosity. In their paper~\cite{1986PhRvA..34.4163T}, Tanaka and Ichimaru derived a form of the collision operator for an OCP including the static local field correction (LFC) $G(k)$ from their earlier theory~\cite{1985PhRvA..32.1768I}. The LFC is defined within a density response formalism by expressing the density-density response in terms of a screened response and LFCs. 
%This amounts to an effective potential between particles produced by a density fluctuation 
This relates the linear response potential surrounding particles to the density fluctuations $\delta\rho$ via, $\Phi(\vc{k},\omega)=Z^2v(k)[1-G(k,\omega)]\delta\rho$ and provides a modified linear dielectric response function $\epsilon(\vc{k},\omega)=1-v(k)[1-G(k)]\chi^{(0)}(\vc{k},\omega)$ where $v(k)=4\pi e^2/k^2$ is the bare Coulomb potential, $\chi^{(0)}(\vc{k},\omega)=-\int d^3p\vc{k}\cdot(\partial f/\partial\vc{p})/(\omega-\vc{k}\cdot\vc{v})$, and $f$ is the distribution function as a function of the momentum $\vc{p}$. With these modifications, their collision operator is
\begin{eqnarray}
C_{\textrm{TI}}(f,f)=m\pi \int\frac{d^3k}{(2\pi)^3}\vc{k}\cdot\frac{\partial}{\partial\vc{p}}\int d^3p'\frac{v^2(k)[1-G(k)]}{|\epsilon(\vc{k},\vc{k\cdot p}/m)|^2}\nonumber\\
\times\delta[\vc{k}\cdot(\vc{p}-\vc{p}')]\vc{k}\cdot\bigg(\frac{\partial }{\partial \vc{p}}-\frac{\partial }{\partial \vc{p}'}\bigg) f(\vc{p})f(\vc{p}') \ \ \ \ 
\end{eqnarray}
and the corresponding generalized Coulomb logarithm relevant to thermal conductivity is
\begin{equation}\label{tixi}
\Xi_{\textrm{TI}}=\frac{2}{\sqrt{\pi}}\int_0^\infty dk \frac{1-G(k)}{k}\int_0^\infty dz \frac{e^{-z^2}}{|\epsilon(k,kv_Tz)|^2}.
\end{equation}
Here, $e$ and $m$ are the particle charge and mass and $v_T=\sqrt{2T/m}$ is the thermal speed. The theory requires calculation of the LFC. To do this, they solve the Ornstein-Zernike equation for the direct correlation function $c(k)$ by using the hypernetted chain (HNC) closure which gives the system of equations~\cite{hansen2006theory}
\begin{eqnarray}
g(\vc{r})=\exp\big[-v(\vc{r})/k_BT+h(\vc{r})-c(\vc{r})+b(\vc{r})\big]\label{hnc1}\\
\hat{h}(\vc{k})=\hat{c}(\vc{k})[1+n\hat{h}(\vc{k})]\label{hnc2}.
\end{eqnarray}
Here, $h(\vc{r})=g(\vc{r})-1$ is the pair correlation function, $b(\vc{r})$ is the bridge function that can be modeled based on a fit to MD data for the OCP\cite{1986PhRvA..34.4163T}, and quantities $\hat{h}$ and $\hat{c}$ are Fourier transforms of $h$ and $c$. The solution for $\hat{c}(\vc{k})$ gives the LFC as $G(\vc{k})=1+k_BTc(\vc{k})/v(\vc{k})$.

The thermal conductivity for the Tanaka-Ichimaru model is compared with the MD data in Fig.~\ref{figTheory}. The prediction asymptotes to the LS theory at small gamma and follows the trend of the MD data for $\Gamma<7$, although it overestimates the value of the MD data by $\sim50\%$ around $\Gamma=1$. At $\Gamma\sim20$ the Coulomb logarithm transitions from a positive to a negative value, so the predicted thermal conductivity diverges. 
This behavior is also observed when applying this model to shear viscosity\cite{2014PhRvE..90c3105D}.

\subsection{Effective Potential Theory}

Like the Tanaka-Ichimaru theory, EPT attempts to extend traditional plasma theory into the strongly coupled regime~\cite{2013PhRvL.110w5001B}. 
%It aims to do this by calculating an effective potential which accounts for many particle interactions and correlation, allowing the binary collision picture to be left intact. 
The concept underlying the theory is that binary collisions do not occur in isolation. 
Rather, surrounding particles influence the effective force through which these collisions occur. 
By taking the two colliding particles at fixed positions, and canonically averaging over the rest of the plasma at equilibrium, the associated interaction potential is the potential of mean force $w(r)$, which is related to the radial distribution function via $g(r)=\exp(-ew(r)/k_BT)$\cite{hill2012introduction}. EPT is based on a Boltzmann-type kinetic theory, but where the interaction potential is taken to be the potential of mean force. 
Recently, this mean force proposition has been formalized by showing that the kinetic theory can be derived by expanding the BBGKY hierarchy in terms of an expansion parameter, associated with perturbations of the distribution function about equilibrium\cite{doi:10.1063/1.5095655}.
Here, the pair distribution function $g(r)$ is calculated via the Ornstein-Zernike equation with HNC closure given in Eqs.~(\ref{hnc1}) and (\ref{hnc2}). 
%Once $g(r)$ is obtained, the effective interaction potential is given by the equilibrium relationship $g(r)=\exp(-e\phi/k_BT)$ with $\phi$ being the potential of mean force on a particle in the plasma. 

\begin{figure}
\includegraphics[width=\columnwidth]{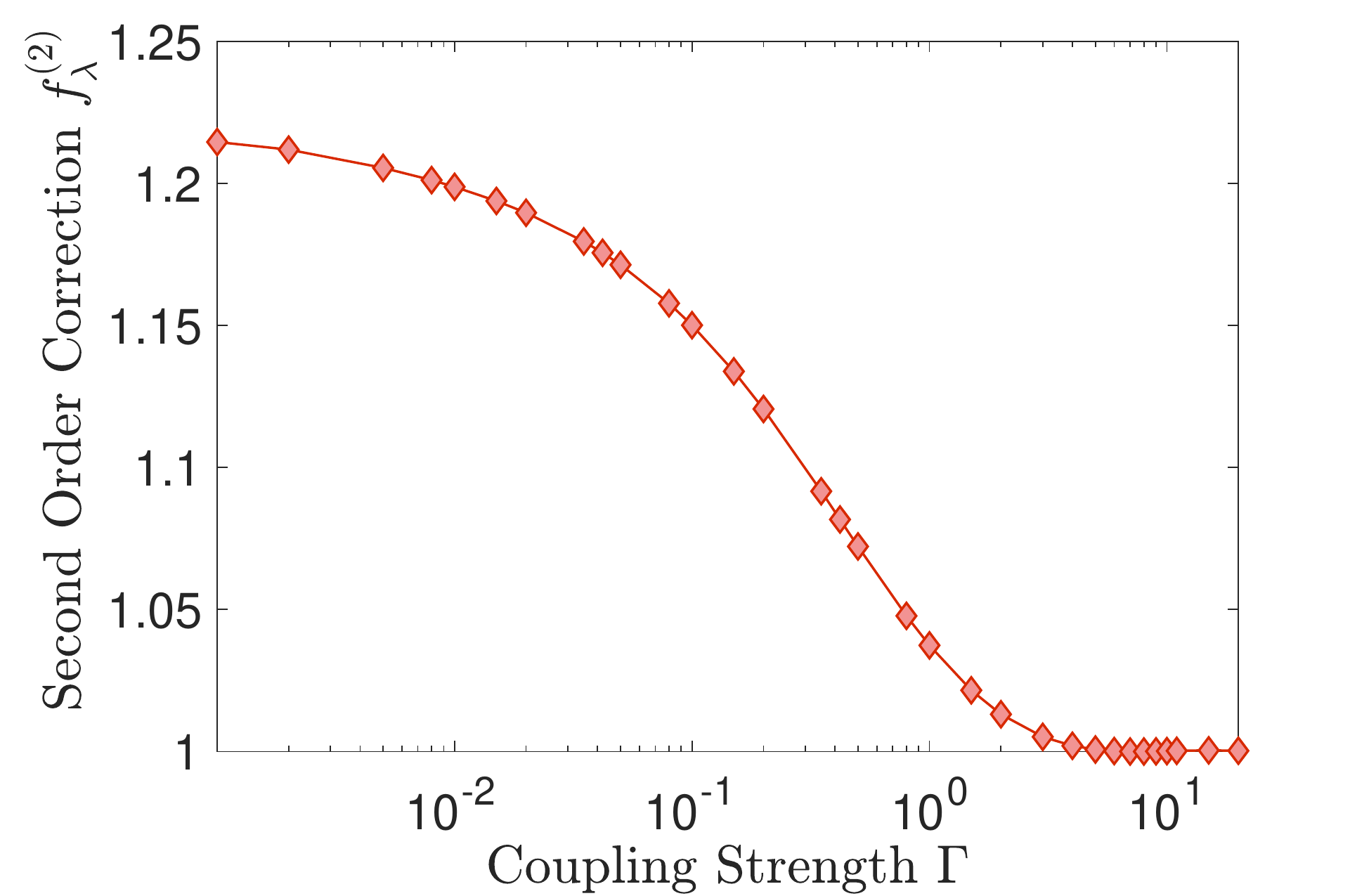}
\caption{\label{l1l2} The second order Chapman Enskog correction for EPT. }
\end{figure}

Once the effective potential is obtained, collision cross sections are calculated for use in the calculation of transport coefficients via the Chapman-Enskog method\cite{ferziger1972mathematical,1970mtnu.book.....C}. 
The $l$th momentum scattering cross section is 
\begin{equation}
\bar{\sigma}^{(l)} = 2\pi \int_0^\infty db\, b [ 1 - \cos^l (\pi - 2 \Theta)] ,
\end{equation}
where
\begin{equation}\label{Th}
\Theta = b\int_{r_o}^{\infty}dr r^{-2}\bigg[1-\frac{b^2}{r^2}-\frac{4w(r)}{mu^2}\bigg]^{-1/2}
\end{equation}
is the scattering angle after a binary collision via the potential $\phi$. 
Here, $r_o$ is the distance of closest approach, which is determined from the largest root of the denominator in Eq.~(\ref{Th}), $u = |\vc{v} - \vc{v}^\prime|$ is the magnitude of the difference of velocities of the colliding particles, and $b$ is the impact parameter. 
%Other quantities needing definition are the reference cross section $\sigma_o=\pi e^2/mv_T^4$, the distance of closest approach $r_o$, the magnitude of the difference in velocities of the colliding particles $u$, and the impact parameter $b$. 
In the Chapman-Enskog theory, transport coefficients can be written in terms of collision integrals $\Omega^{(l,k)}$\cite{1970mtnu.book.....C,ferziger1972mathematical}, which can be recast in terms of generalized Coulomb logarithms $\Xi^{(l,k)}=\sqrt{\pi}\Omega^{(l,k)}/v_T$, where
\begin{equation}\label{xi}
\Xi^{(l,k)} = \frac{1}{2} \int_0^\infty d\xi\, \xi^{2k+3} e^{-\xi^2} (\bar{\sigma}^{(l)} /\sigma_o)
\end{equation}
where $\sigma_o=\pi e^2/mv_T^4$ is a reference cross section. 
The thermal conductivity from the second order Chapman-Enskog solution is\cite{ferziger1972mathematical}
\begin{equation}\label{lept}
\lambda^*_{\textrm{EPT}}  = \frac{25}{4} \frac{\sqrt{\pi/3}}{\Gamma^{5/2} \Xi^{(2,2)}}f_\lambda^{(2)}
\end{equation}
where
\begin{equation}
f_\lambda^{(2)}=1+\frac{49(\Xi^{(2,2)})^2-28\Xi^{(2,2)}\Xi^{(2,3)}+4(\Xi^{(2,3)})^2}{28(\Xi^{(2,2)})^2+4\Xi^{(2,2)}\Xi^{(2,4)}-4(\Xi^{(2,3)})^2}
\end{equation}
is the second order correction. Although the second order correction (shown in Fig.~\ref{l1l2}) is included, it is small at strong coupling. This reflects the fact that highly collisional strongly coupled systems tend to have velocity distribution functions closer to equilibrium. 

Figure~\ref{figTheory} shows the result of EPT obtained by applying Eq.~(\ref{lept}) in Eq.~(\ref{lstar}).
 %with $\Xi^{(2,2)}$ calculated using the effective potential. 
Like the Tanaka-Ichimaru model, the EPT theory slightly overshoots the MD data between $\Gamma\approx0.5-7$ but still captures the scaling with $\Gamma$. The model fails to capture the scaling for $\Gamma \gtrsim 7$ where the potential and virial terms become important. Previous calculations of other transport coefficients required an accounting for the Coulomb hole in the pair distribution function at strong coupling to eliminate the overshoot of the MD data~\cite{2015PhRvE..91f3107B}. %In their work, Baalrud and Daligualt 
This work added additional physics to the model via a generalized Enskog kinetic theory that describes an increased collision frequency $\chi$ resulting from the reduced volume in which repulsively-interacting charged particles can occupy. Figure~\ref{figTheory} shows that the thermal conductivity with this correction calculated via the method in Ref.~\onlinecite{2015PhRvE..91f3107B},
\begin{equation}
\lambda^*_{\textrm{EPT-Enskog}}=\frac{\lambda^*_{\textrm{EPT}}}{\chi},
\end{equation}
is in good agreement with the MD data over the range $\Gamma \lesssim 7$.

The range of $\Gamma$ over which there is agreement is similar to that of the shear viscosity, but lower than the self diffusion for which EPT is accurate for $\Gamma\lesssim 40$~\cite{2015PhRvE..91f3107B}. 
The difference in range of agreement for these three transport coefficients is due to the components which go into their calculation. For diffusion, the Green-Kubo relation only requires the velocity ACF. The diffusion coefficient only depends on the mass transport and does not have a component solely due to the potential. 
% that only leads to modifications that change the collision rate and change in velocity of the individual particles. 
In contrast, the Green-Kubo relations for the shear viscosity and thermal conductivity both involve fluxes which have potential or potential and virial parts in addition to the parts related to kinetic energy transport. Agreement with MD data is only expected when the transport coefficient is determined by the kinetic component. 
Indeed, this can be seen in Fig.~\ref{figTheory}, where EPT is found to be in much better agreement with the kinetic component of the thermal conductivity to a much larger coupling strength than it is with the total thermal conductivity.

\section{Summary}
We have presented calculations of thermal conductivity of the OCP from equilibrium MD simulations and have addressed the numerical difficulty which has caused previous disagreement between equilibrium and non-equilibrium MD calculations. Analysis of ACFs reveal that noise due to the finite extent of the heat flux time series can dominate the value of the thermal conductivity before the cumulative integral has converged to its final value. To avoid this source of error which has plagued previous equilibrium MD simulations, the value of the cumulative integral for long $6\times10^5\omega_p^{-1}$ time series were averaged over the interval [100$\omega_p^{-1}$, 150$\omega_p^{-1}$] and then averaged for 6 characteristic time series. The values obtained are in good agreement with previous non-equilibrium MD data for $2\le\Gamma\le180$. 

New MD data in the range $0.1\le\Gamma\le2$ was also presented. This facilitates the comparison of the thermal conductivity calculated with generalized Coulomb logarithms from different theoretical models with MD data. The newly obtained low $\Gamma$ data was found to be in good agreement with the Landau-Spitzer theory when $\Gamma \lesssim 0.3$, providing the first computational test of this transport coefficient. At stronger coupling, a comparison of EPT and Tanaka-Ichimaru theories with MD data demonstrates that the effect of particle correlations needs to be accounted for to reproduce the trend of the simulated thermal conductivity for $0.3<\Gamma<7$. The comparison also demonstrates that the treatment of strong coupling in the Lee-More model, which amounts to replacing the Debye length with the ion sphere radius, is inadequate. This realization may be important for simulations of high energy density experiments in which this model is commonly used. Finally, this study finds that none of the models can accurately predict the thermal conductivity for $\Gamma \gtrsim 10$. This is an indication of the limit of the effective binary collision picture. At stronger coupling, simulations indicate that the virial, potential, and cross components dominate, indicating that multiparticle dynamics determines the transport. The comparison with MD data demonstrates the need for theories which can properly describe transport at stronger coupling. 

\section*{Acknowledgements}
This work was supported by the U.S. Department of Energy, Office of Fusion Energy Sciences, under Award No.~DE-SC0016159.

\bibliography{lambda}

\end{document}